\documentclass[usenatbib]{mn2e}
\usepackage{graphicx}

\def\gax{\mathrel{\raise.3ex\hbox{$>$}\mkern-14mu\lower0.6ex\hbox{$\sim$}}}
\def\lax{\mathrel{\raise.3ex\hbox{$<$}\mkern-14mu\lower0.6ex\hbox{$\sim$}}}
\def\gtorder{\mathrel{\raise.3ex\hbox{$>$}\mkern-14mu
             \lower0.6ex\hbox{$\sim$}}}
\def\ltorder{\mathrel{\raise.3ex\hbox{$<$}\mkern-14mu
             \lower0.6ex\hbox{$\sim$}}}

\begin{document}

\title [Gaia and RSG Convection]
   {A Non-Detection of Red Supergiant Convection in Gaia}

\author[C.~S. Kochanek]{ 
    C.~S. Kochanek$^{1,2}$ 
    \\
  $^{1}$ Department of Astronomy, The Ohio State University, 140 West 18th Avenue, Columbus OH 43210 \\
  $^{2}$ Center for Cosmology and AstroParticle Physics, The Ohio State University,
    191 W. Woodruff Avenue, Columbus OH 43210 \\
   }

\maketitle

\begin{abstract}
Large scale surface convection on red supergiants (RSGs) can lead to shifts in the
photocenter of the star which might be measured by Gaia and used as a new probe of
the surface dynamics of these rare but important stars.  Unlike brightness variations,
photocenter motions would provide information on the physical scale of the convective
cells. The signal would be that RSGs show an excess astrometric noise 
at the level of a few percent of the stellar radius.
Unfortunately, we find that the excess astrometric noise level of Gaia EDR3 is roughly 
an order of magnitude too large to detect the predicted motions and that
RSGs have excess astrometric noise indistinguishable from other stars of similar
magnitude and parallax.  The typical excess astrometric noise steadily decreases 
with $G$ magnitude (for $G<11$~mag), so it is crucial to compare stars of similar
brightness. 
\end{abstract}


\section{Introduction}

Most of the stars exploding as supernovae are red supergiants (RSGs) producing
Type~IIP/IIL supernovae (e.g., \citealt{Smartt2009}, \citealt{Li2011}).  Models of
the light curves of these supernovae have difficulty matching the data, leading
to invocations of large scale mass loss just prior to the explosion (e.g.,
\citealt{Morozova2018}, \citealt{Morozova2020}).  However, direct observations
of the pre-supernova variability of RSGs find that they vary no more than
any other RSG (\citealt{Kochanek2017}, \citealt{Johnson2018}), which would
seem to rule this possibility out.

It is, however, possible that the problem lies in the structure of the RSG
progenitors used for the light curve models.  The atmospheres of RSGs are 
so unstable that stellar evolution models generally apply an artificial external
pressure to stabilize them (e.g., \citealt{Heger2000}), giving the progenitor
model a sharp outer edge.  In real stars, however, the atmospheres are dynamical
with large scale convective flows that are directly observed in the case
of Betelgeuse (e.g.,
\citealt{Haubois2009}, \citealt{Lopez2022}).  In simulations, these surface
flows lead to a very extended, dynamic atmosphere (e.g., \citealt{Arroyo2015},
\citealt{Goldberg2022}) which is not present in supernova progenitor models.

\citet{Chiavassa2011} noted that the brightness changes associated with the
convective cells would cause the photocenter of an RSG to wander and that 
this might be detectable with Gaia (\citealt{Gaia2016}, \citealt{Gaia2021}).  
They also argued that there was evidence
for the effect in the Hipparcos (\citealt{ESA1997}, \citealt{vanLeeuwen2007})
observations of Betelgeuse. In \citet{Chiavassa2018} and \citet{Chiavassa2022},
they argue for detections of the effect in Gaia observations for Asymptotic
Giant Branch (AGB) stars and RSGs, respectively.  The discussion of RSGs
appeared as we were doing this analysis and will be discussed.  

The expected signal is a fluctuation in the photocenter of the star by
$\sigma_R = \epsilon R_*$ where $R_*$ is the radius of the star and
$\epsilon \simeq 0.023 \pm 0.015$ for the six models in \cite{Chiavassa2022}.
The shift in the angular position is then 
$\sigma_\theta=\sigma_R/d=\epsilon R_*/d=\epsilon\theta_*$ where
$d$ is the distance to the star.  Note that $\sigma_\theta$ is independent
of distance if the radius of the star is derived from the observed spectral
energy distribution: $\theta_* = R_*/d \propto L_*^{1/2}/d \propto  F_*^{1/2}$ 
for luminosity $L_*$ and flux $F_*$.   
The parameter in the Gaia astrometric solutions which is directly sensitive
to the photocenter wandering is the astrometric excess noise $\sigma_e$
(see \citealt{Lindegren2021}).  This is the quantity which must be added
in quadrature to the known sources of noise in the astrometric measurements
to produce a statistically good
astrometric fit. The photocenter wander $\sigma_\theta$ is simply one
term contributing in quadrature to $\sigma_e$.  Given a sample of RSGs spanning
a range of angular sizes $\theta_*$, the expected signal is an excess
noise $\sigma_e \propto \theta_*$ with an amplitude of $\sigma_\theta$.
If the photocenter wander does not dominate the excess noise,
then the effect is undetectable absent a quantitative knowledge of the
other effects that dominate the excess noise.
In \S2 we describe our search for the signal and how we reach
the conclusion that the signal is presently undetectable.  
As noted above, \cite{Chiavassa2022} came to the opposite conclusion
using stars in the $\chi$~Per cluster as we were working on this study. 
In \S3 we show that the conclusion of \S2 also holds for these stars.
We discuss future prospects in \S4.

\begin{figure}
\includegraphics[width=0.50\textwidth]{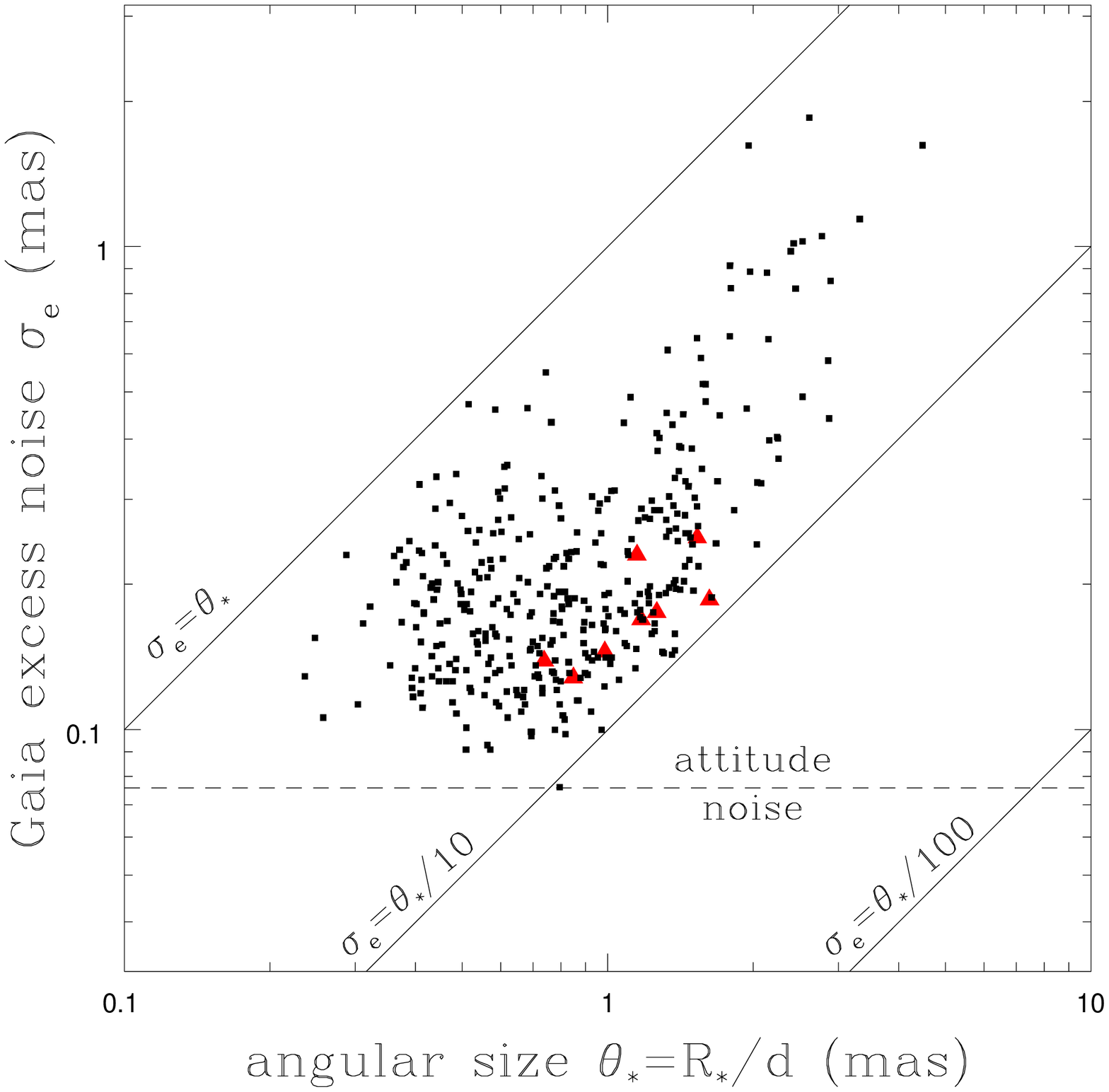}
\caption{
  Gaia EDR3 astrometric excess noise $\sigma_e$ as a function of the angular size $\theta_*$
  of the star.  The black points are the group A and B RSG candidates from 
  \protect\cite{Messineo2019} and the red triangles are the $\chi$~Per RSGs
  used by \protect\cite{Chiavassa2022}.  The lines show
  where $\sigma_e=\theta_*$, $\theta_*/10$ and $\theta_*/100$.  The astrometric shifts due
  to convection should be a few percent (\citealt{Chiavassa2022}).
  The horizontal dashed line is the median spacecraft attitude noise contribution
  to the excess astrometric noise (\protect\citealt{Lindegren2021}). 
  }
\label{fig:results1}
\end{figure}

\section{Searching for the Signal}

We used the \cite{Messineo2019} catalog of RSGs. The catalog is based on modeling
the spectral energy distributions of red stars with Gaia DR2 parallaxes.  The initial
catalog is heavily contaminated by asymptotic giant branch (AGB) stars
so we focus on their groups A and B which should be dominated by RSGs.
We convert their estimates of the luminosity and temperature to radius
$R_*$ and then to the angular size $\theta_*$ using the same 
\cite{BailerJones2018} distances used by \cite{Messineo2019}.  Because
$\theta_*$ is independent of distance, it is important to use the same
distance as was used to estimate $L_*$. We also only used stars with
parallax $\varpi > 0.1$~mas and $G<11$~mag.  We use the Gaia EDR3 (\citealt{Gaia2021},
\citealt{Lindegren2021})
astrometric data for the actual analysis. This left us with
360 stars. Fig.~\ref{fig:results1} shows
the results, with a remarkably good linear correlation between 
$\theta_*$ and $\sigma_e$ as expected for photocenter fluctuations.
While this initially led to a frisson of excitement, there is a problem --    
the amplitude is roughly an order of magnitude too large, with 
$\sigma_e \sim \theta_*/4$ ($\langle \log(\sigma_e/\theta_*)\rangle =-0.6$ with
a scatter of $0.2$~dex) instead of $\sigma_e \sim 0.02 \theta_*$.  

The excess astrometric noise is actually the quadrature sum
of a time dependent excess noise associated with the satellite and an
excess noise associated with the particular source (\citealt{Lindegren2021}).
Obviously, this analysis should be done with the excess noise
associated with the individual source, but there seems to be no way
to access the two contributions separately in the existing interfaces.   
\cite{Lindegren2021} reports that the median excess attitude noise
is 0.076~mas, and this noise level is also shown in Fig.~\ref{fig:results1}.
If this is the dominant source of spacecraft noise, then the
total excess noise is dominated by the source contribution and
subtracting the $0.076$~mas in quadrature from $\sigma_e$ does not markedly 
change the character of Fig.~\ref{fig:results1}.

We can also examine whether the excess noise seen for the RSGs
is any different from that of other stars.  We do this by building a comparison
sample of stars with the similar G magnitudes and
parallaxes for each RSG.  We selected stars 
within $\Delta G = 0.25$~mag in magnitude, have parallaxes between $0.8\varpi$
and $1.2\varpi$ of the RSG parallax $\varpi$ 
and excluding the RSGs themselves.  We sorted each 
RSGs comparison stars by their astrometric excess noise and then
found the rank of the RSG relative to the comparison sample.
A rank of $0$ ($1$) means the excess noise of the RSG was
less (greater) than that of all the comparison stars and 
a rank of $0.5$ means it is at the median.    

RSGs are sufficiently rare that there is should be no problem from 
contamination by unrecognized RSGs, but we might
worry about contamination by (pulsating) AGB stars since 
\cite{Chiavassa2018} argue for a detection of the photocenter
wobble in AGB stars. As a 
check for any such effects, we subdivided the comparison
samples into blue ($B_P-R_P<1$) and red ($B_P-R_P>1$) stars.
The blue stars should generally be much smaller, non-pulsating main sequence
(MS) stars with no photocenter shifts due to convection (simply
because they are much smaller in radius if nothing else).
We required the comparison sample to have at least 10 stars,
which left us with 346, 326, and 334 stars for the all,
blue, and red comparison samples.

\begin{figure}
\includegraphics[width=0.50\textwidth]{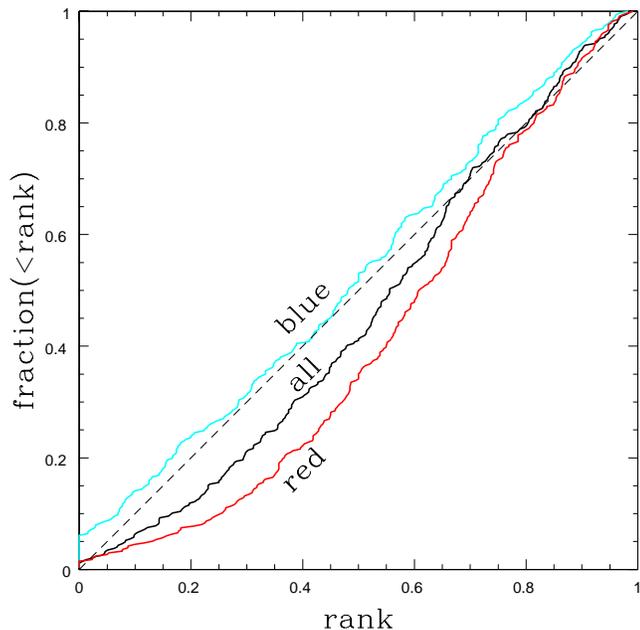}
\caption{
  Distribution of the Gaia excess astrometric noise ranks of the group A and B RSG candidates from 
  \protect\cite{Messineo2019} relative to their comparison samples of all (black), red (red) and
  blue (cyan) stars.  If the RSG ranks were randomly sampling the excess noise distribution of the
  comparison sample, they would follow the dashed line.
  }
\label{fig:results2}
\end{figure}

Fig.~\ref{fig:results2} shows the distributions of the RSGs by
rank.  If the ranks of the RSGs were randomly chosen from their
comparison samples, then their rank distribution should simply
be straight line, and we see this is very close to what we see
for the blue stars.  The RSGs tend to have higher ranks in the
red star comparison samples (i.e. the red star curve is below
the dashed line), but only for low and intermediate ranks.  The
fraction of RSGs at the highest ranks (most excess noise) are
again very similar to the comparison sample.  Compared to red
stars, there are fewer RSGs with low excess noise, more with
intermediate excess noise, and comparable numbers with high
excess noise.  Since the full sample is a combination of the
red and blue samples, the rank distribution for all stars is
intermediate to the red and blue distributions.  Overall, the
astrometric noise of the RSGs is not very different from the
noise seen for stars of similar magnitude and parallax. 

Between the high amplitudes and the similarity to other stars,
it is clear that the astrometric shifts due to convection are
not presently detected.    
Since the RSGs are of similar sizes 
and temperatures, the $\log \theta_*$ axis in Fig.~\ref{fig:results1} 
is essentially the same as the logarithm of the flux (i.e., the G 
magnitude).  This suggests that the correlation in
Fig.~\ref{fig:results1} is really showing
that the excess noise depends on magnitude and the correlation
with magnitude just happens to mimic the linear correlation expected 
for the photocenter motions due to convection.  Exploring this 
leads to an explanation of the \cite{Chiavassa2022} results.  

\begin{figure}
\includegraphics[width=0.50\textwidth]{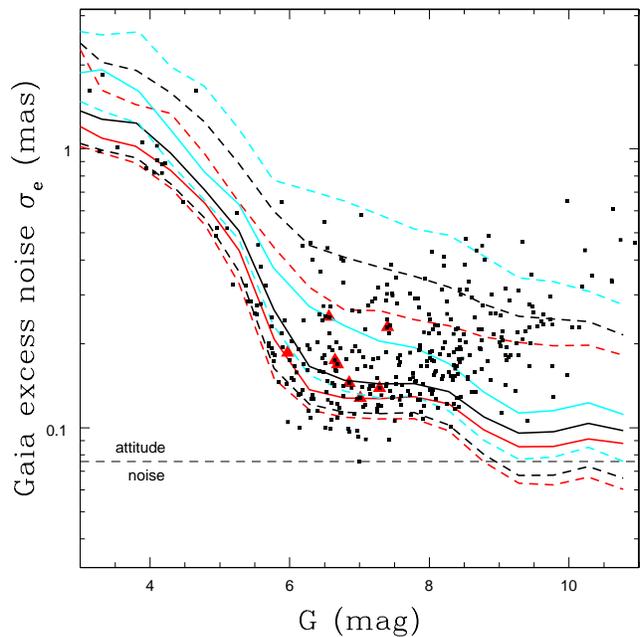}
\includegraphics[width=0.50\textwidth]{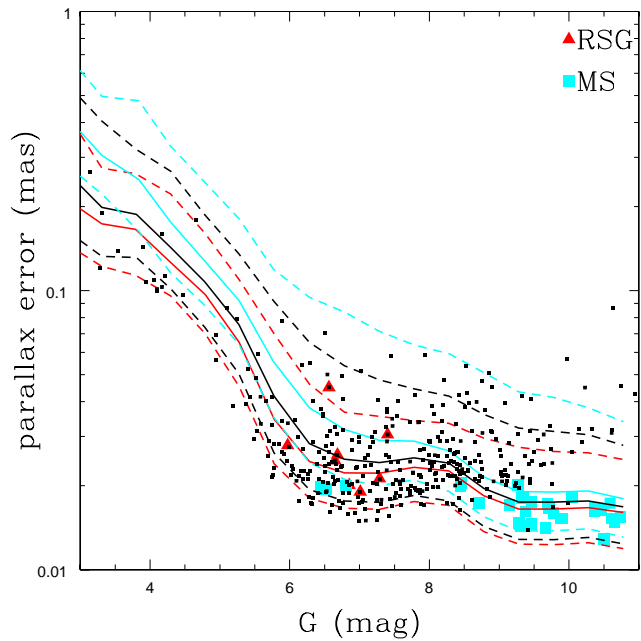}
\caption{
  Gaia EDR3 astrometric excess noise $\sigma_e$ (top) and parallax error $\sigma_\varpi$ (bottom)
  as a function of apparent G magnitude.  Solid (dashed) lines are the median ($1\sigma$ range)
  for all (black), blue (cyan, $B_P-R_P<1$~mag) and red (red, $B_P-R_P>1$~mag) stars.
  The small black squares are the group A and B RSG candidates from \protect\cite{Messineo2019},
  the red triangles and cyan squares are the RSGs and MS stars in $\chi$~Per from 
  \protect\cite{Chiavassa2022}.  The MS stars are not shown in the top panel.
  The horizontal dashed line in the top panel is the median spacecraft attitude noise contribution
  to the excess astrometric noise (\protect\citealt{Lindegren2021}). 
  }
\label{fig:results3}
\end{figure}

\section{The Magnitude of the Problem}

As we were working through the discussion in \S2, \cite{Chiavassa2022}
appeared with the claim of a detection at the expected amplitude.  
Needless to say, this was puzzling given the results of \S2.  \cite{Chiavassa2022} examined
the parallax measurement errors for 8 RSGs and 30 MS stars in the
$\chi$~Per cluster, finding that the RSGs had parallax errors larger
than the MS stars roughly by the expected amplitude of the 
astrometric jitter from convection at the distance to
$\chi$~Per ($2.3$~kpc). They interpret this as a detection of
the convectively driven motions. 

Fig.~\ref{fig:results1} also includes the 8 $\chi$~Per RSGs,
and they have astrometric excess noises consistent with the
sample we considered (although they seem not to be included
in the \cite{Messineo2019} sample).  So like that sample,
the excess noise is an order of magnitude too large to be
due to convection.  The effect on the parallax errors is 
much smaller simply because the parallax and its errors 
come from a fit to all of the astrometric measurements.  The
RSGs in \S2 had an average of $N=423$ good astrometric 
measurements 
({\tt astrometric\_n\_good\_al}), so the contribution 
of the excess noise to
the parallax error is $\sigma_\varpi^2 \sim \sigma_e^2/N$
where there is a dimensionless coefficient that depends on
the sensitivity of the parallax to measurements and the 
level of parameter degeneracies.  With $N^{1/2} \simeq 20$,
the numbers of measurements are of the right order of 
magnitude to convert the $\sim R_*/4$ excess noise into
a $\sim R_*/80$ parallax error that is comparable to the
expected signal.  But, the convective motion signal is
a direct contribution to the excess noise and then gets
reduced by averaging as a contribution to astrometric 
variables like the parallax.  

At the end of \S2, our hypothesis for explaining Fig.~\ref{fig:results1}
was that the excess noise must depend on magnitude.  To explore this,
we selected all Gaia EDR3 stars with
$G<8$~mag and then random samples of stars down to $G=11$~mag.  
We also restricted the stars to have $\varpi >0.1$~mas and
$\varpi > 4 \sigma_\varpi$.  For
bins $0.5$~mag wide we computed the median excess astrometric error
and parallax error along with their symmetric $1\sigma$ ranges.  We
computed the statistics for all stars, blue stars ($B_P-R_P<1$~mag)
and red stars ($B_P-R_P>1$~mag).  These are shown in Fig.~\ref{fig:results3}.
The errors show a systematic decrease with increasing magnitude
for both quantities and all three ways of dividing the sample.
\cite{Lindegren2021} includes the lower panel over a broader
magnitude range, and the typical parallax uncertainties remain
roughly constant to $G \sim 13$ and then begin to increase for
fainter sources.

These trends explain the correlation
in Fig.~\ref{fig:results1} as expected.  The $\theta_*$ axis is basically just
flux where the largest stars have $G \simeq 2.5$ and the smallest
stars have $G \simeq 11$, so Fig.~\ref{fig:results1} is simply a
recasting of the trends in Fig.~\ref{fig:results3}.  Fig.~\ref{fig:results3}
also include the \cite{Messineo2019} RSGs and they largely follow
the trends for all other stars, consistent with the rank distribution
in Fig.~\ref{fig:results2}.  The faintest RSGs do begin to
lie above the trends, and this turns out to be a consequence of
selecting stars only in magnitude and not also in parallax.   This
was done so that there would be enough stars to do these statistics
at bright magnitudes.  

Fig.~\ref{fig:results3} also include the $\chi$~Per RSGs and MS
stars from \cite{Chiavassa2022}.  For their magnitudes, the $\chi$~Per
stars have relatively low excess noise and parallax error.  This is
likely due to differences between the number and pattern of Gaia scans
of $\chi$~Per as compared to an average location.  However, most of
the MS stars are fainter than the RSGs, so they have less excess noise
and smaller parallax errors simply because both quantities decrease
as you examine fainter stars over this magnitude range.  The RSGs are noisier
than the two brightest MS stars, but not to a degree that is statistically
significant compared to red stars of similar brightness.

\section{Discussion}

It is a remarkable feat that Gaia can measure the angular positions of
these RSGs to a small fraction of the stellar radii.  Unfortunately,
the sources of excess astrometric noise from other sources are presently
too large to detect the predicted photocenter variability created by
convection on the surfaces of RSGs.  The observed excess noise is 
roughly an order of magnitude too large (Fig.~\ref{fig:results1}).  
If we build a comparison sample of stars for each RSG with similar magnitudes 
and parallaxes, the RSGs do not stand out for having distinctive amounts
of excess astrometric noise (Fig.~\ref{fig:results2} and \ref{fig:results3}). 

\cite{Chiavassa2022} argue for a detection based on the parallax error
differences between RSG and MS stars in the $\chi$~Per cluster.  However,
most of the difference between the RSG and MS stars is simply due to the magnitude
dependence of the errors -- for the magnitude range in question, fainter
stars have smaller parallax errors (see \citealt{Lindegren2021}).
More generally, both the excess noise and parallax errors of the $\chi$~Per
RSGs are typical of stars of their magnitude.  In fact, they are modestly
smaller than average, presumably because of differences between the average
Gaia coverage and the detailed coverage of $\chi$~Per.  That the difference
in parallax errors is comparable to the expected photocenter wander is
simply a coincidence.  

The detection of the photocenter variability due to convection remains
an interesting physical prospect, but it requires a reduction in the
Gaia excess astrometric noise from other sources by over an order of
magnitude.  As noted earlier, the excess noise is a quadrature sum
of contributions from the spacecraft and the source where for the stars
we consider, the reported median space craft contribution ($0.076$~mas,
\citealt{Lindegren2021}) appears to be sub-dominant. Nonetheless, it
would be useful to have the estimated source noise contribution separate from
the spacecraft noise contribution.  Note, however, that the excess noise
associated with the spacecraft is frequently going to be larger than the 
expected contribution of the photocenter motions to the excess noise

\section*{Acknowledgments}

The author thanks T.~Thompson for suggesting this as a question to look into.
CSK is supported by NSF grants AST-1908570 and AST-1814440.  
This work has made use of data from the European Space Agency (ESA) mission
{\it Gaia} (https://www.cosmos.esa.int/gaia), processed by the {\it Gaia}
Data Processing and Analysis Consortium (DPAC,
https://www.cosmos.esa.int/web/gaia/dpac/consortium). Funding for the DPAC
has been provided by national institutions, in particular the institutions
participating in the {\it Gaia} Multilateral Agreement.

\section*{Data Availability Statement}

All data used in this paper are publicly available.

\end{document}